# The Cure: Making a game of gene selection for breast cancer survival prediction

Benjamin M. Good[*,1], Salvatore Loguercio[1], Obi L. Griffith[2], Max Nanis[1] , Chunlei Wu[1], and Andrew I. Su[1]

1. Department of Molecular and Experimental Medicine, The Scripps Research Institute,

10550 North Torrey Pines Road, La Jolla, CA

2. Department of Medicine, Washington University School of Medicine

4444 Forest Park Ave, St. Louis, MO

## ABSTRACT

**Motivation:** Molecular signatures for predicting breast cancer prognosis could greatly improve care through personalization of treatment. Computational analyses of genome-wide expression datasets have identified such signatures, but these signatures leave much to be desired in terms of accuracy, reproducibility and biological interpretability. Methods that take advantage of structured prior knowledge (e.g. protein interaction networks) show promise in helping to define better signatures but most knowledge remains unstructured.

Crowdsourcing via scientific discovery games is an emerging methodology that has the potential to tap into human intelligence at scales and in modes previously unheard of. Here, we developed and evaluated a game called "The Cure" on the task of gene selection for breast cancer survival prediction. Our central hypothesis was that knowledge linking expression patterns of specific genes to breast cancer outcomes could be captured from game players. We envisioned capturing knowledge both from the player's prior experience and from their ability to interpret text related to candidate genes presented to them in the context of the game.

**Results:** Between its launch in Sept. 2012 and Sept. 2013, The Cure attracted more than 1,000 registered players who collectively played nearly 10,000 games. Gene sets assembled through aggregation of the collected data clearly demonstrated the accumulation of relevant 'expert' knowledge. In terms of predictive accuracy, these gene sets provided comparable performance to gene sets generated using other methods including those used in commercial tests. The Cure is available at http://genegames.org/cure/

## 1    INTRODUCTION

Breast cancer is the most common form of cancer in women (Bray, et al., 2013). It has been studied extensively with genomic technologies, with many attempts to devise molecular predictors of clinical outcomes (Griffith, et al., 2013; Margolin, et al., 2013; van 't Veer, et al., 2002) and drug response (Daemen, et al., 2013). If successful, tests derived from these predictors would pave the way towards personalized therapy and better care. While much progress has been made, including several commercially available tests (Ross, et al., 2008), molecular predictors consistently show lower than desirable accuracy, degrade in performance in subsequent validation studies, identify different gene sets in every permutation, and often have no discernable biological rationale (Weigelt, et al., 2012).

Any attempt at class prediction based on high throughput (e.g. microarray) data is technically challenging because of the very large number of potential features (Cheng, et al., 2013b). The individual members of optimal feature sets work synergistically, displaying relationships that make the group more useful for prediction as a whole than any individual unit. The space of possible feature combinations is too large to explore exhaustively and, even if it were, the tests available for evaluating feature set quality are not precise. As a result, researchers rely on heuristics and, increasingly, on prior knowledge to identify good feature groups.

Recent gene selection methods are driven by structured prior knowledge in forms such as protein-protein interaction networks (Dutkowski and Ideker, 2011; Winter, et al., 2012), pathway databases (Bild, et al., 2006; Su, et al., 2009) and information gathered from pan-cancer datasets (Cheng, et al., 2013a). These methods guide the search for predictive gene sets towards cohesive groups related to each other and to the predicted phenotype through biological mechanism. In doing so, they have improved the stability of the gene selection process and the biological relevance of the identified signatures. These techniques hint at the potential of strategies that marry a top-down approach based on established knowledge with a bottom-up approach based directly on experimental data, but they have not yet produced substantially greater accuracy than other approaches. This may be due in part to a scarcity of relevant structured knowledge with which to compute.

Since the year 2000, more than 164,000 publications related to breast cancer have been added to PubMed (http://tinyurl.com/brsince2000). Within that body of literature and in the minds of those that have created and consumed it lays a wealth of knowledge relevant to selecting gene sets for survival prediction. Here, we explore a crowdsourcing approach for tapping into that knowledge.

Crowdsourcing processes take tasks traditionally performed by individuals or small groups and reformulate them such that large numbers of people can participate in their completion. There are many instantiations of the crowdsourcing paradigm (Good and Su, 2013), here we focus on just one: games with a purpose (GWAP) (von Ahn and Dabbish, 2008). GWAPs incentivize large-scale work by translating the required labor into elements of games. The games are played for fun, for learning, and to aid in achieving the underlying purpose. Popular GWAPs within biology include Foldit, for protein folding (Cooper, et al., 2010), Phylo for multiple sequence alignment (Kawrykow, et al., 2012) and MalariaSpot for image analysis (Luengo-Oroz, et al., 2012). Here we introduce a GWAP for genomic feature selection called The Cure.

---

[*]To whom correspondence should be addressed.

Our high-level objective is to identify genes that can be used to build improved prognostic predictors for breast cancer. Our hypothesis is that, if aggregated effectively, the collective knowledge, reading, and reasoning ability of a large community could help to identify genes that are useful in constructing robust classifiers but might be hidden from purely data-driven approaches. In striving to achieve that aim, we conducted the study described here to assess the feasibility of the use of an open, online game (The Cure) in capturing pertinent, expert-level biomedical knowledge.

The central questions addressed are:

(1) How many people, of what levels of expertise, would play a game oriented around gene selection for breast cancer survival prediction and why?

(2) Would it be possible to extract a gene ranking from the results of play that reflected biomedical knowledge? That is, could the game act as a portal for expert-level knowledge transfer?

(3) Could the gene ranking captured through the game be used to generate classifiers that perform well in cross-dataset evaluations?

The null hypotheses are that (1) no one would play, (2) the results of their play would not yield discernible biological knowledge, and (3) any gene ranking produced would be no better than random. Below we discuss the design of the game and then present results from one year of open play that shed light on each of the questions posed above.

## 2 GAME DESIGN

The Cure is a web application consisting of the pages: home, login, board selection, game and help. The home page provides information about the project and the game and allows users to either login or create accounts. Users must create an account to play. During account creation, users must select a username and password and have the option of entering an email address and answering three short survey questions: "Most recent academic degree?", "Do you consider yourself knowledgeable about cancer biology?", and "Do you consider yourself a biologist?".

**Training**

When players first register, they are presented with a training stage that must be passed before they enter the main game area. The training stage consists of four 'boards' containing 2 to 4 features common to animals such as "number of legs, breathes air, produces milk, etc.". To complete the level, the player must select the features that can best be used to discriminate between mammals and other classes of animal - before the games automated opponent 'Barney' beats them to it. This task was chosen as a way to introduce the dynamics of the game and to get across the idea of feature selection for classification on a straightforward problem.

**Game**

After training, the player is presented with boards containing 25 different genes (Fig. 1). The objective of each game is to choose a set of 5 genes that produces a better decision tree classifier than that of the automated opponent 'Barney'. The players alternate turns taking a gene card from the board and placing it in their hand, with the human player always going first. Once a card is taken from the board, it cannot be put back and the other player cannot take it. The score for the final 5-card hand determines the winner of the game. Note that each time a board is rendered, the locations of the genes are randomized to prevent bias.

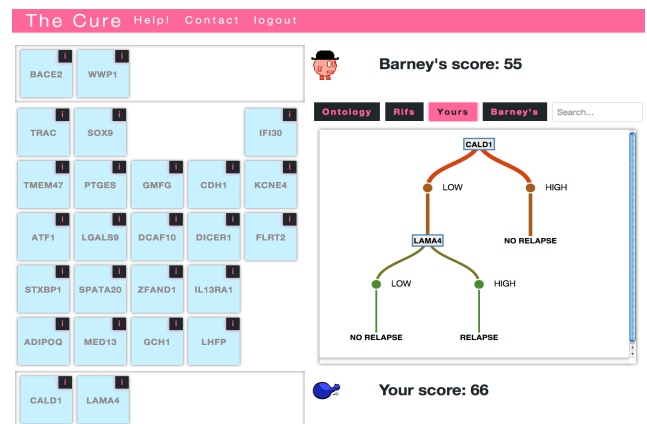

**Fig. 1.** The Cure game. Players alternate turns taking a gene card from the board and adding it to their hand. The tabbed display provides gene annotations ('ontology', 'Rifs') and views of decision trees constructed by the system using the selected genes.

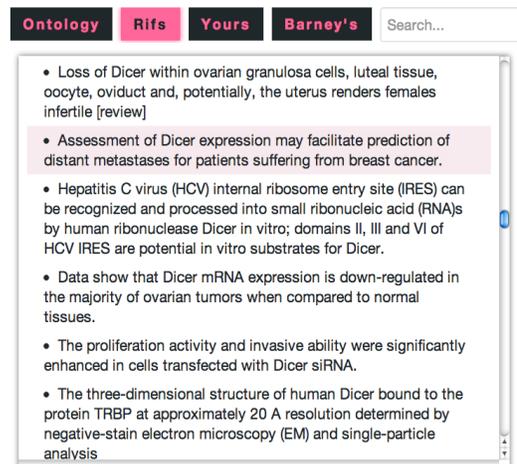

**Fig. 2.** Gene Rifs tab showing information about the Dicer gene.

## Gene annotations

Mousing over each gene provides the player with information including: a summary description from Unigene, Gene Ontology annotations, and related Gene Rifs (Fig. 2). All of the annotations contain hyperlinks that the players can follow for more information. A search interface allows the player to find genes on the board based on the text in their related annotations. Coupled with the player's biological knowledge, this information helps the player make informed guessed about which genes from the board might make the most useful predictors.

## Scoring

Each time a card is added to a player's hand, the game server scores the hand by evaluating the combined predictive performance of the genes it contains. To accomplish this evaluation, the server uses a gene expression dataset containing samples classified with long-term (>10 year) survival status. In each evaluation, the server uses data from just the genes in the player's hand to train and test a decision tree classifier. The score for the hand is the accuracy returned by a cross-validation experiment. In machine learning parlance, this is known as a "wrapper" feature set evaluation scheme (Kohavi and John, 1997). A simplified decision tree created using all of the available training instances but just the selected genes is displayed for the player and their opponent (Fig. 1). If the player loses they are not awarded any points, they may play the board again or select a different board to play. If they win, their score is determined based on the accuracy of their winning tree. Within each round, player scores are cumulative. The more games they win, the higher their score. The player's score is displayed on the board selection page along with its global rank and the current top 10 scores.

## Board selection

Each round of The Cure consists of a collection of 100 different boards for players to choose from (Fig. 3). Each board is composed of a different set of 25 genes (See Methods for board composition strategies). The boards are arranged in loose order of difficulty with the easiest boards occupying the lower numbers. Difficulty is assessed based on an estimation of the predictive power of the complete 25 gene set - the more predictive, the easier the board. The goal of the board selection page is to capture both broad and deep coverage of all the boards (and their corresponding gene sets) by the player community. Once a given board has been completed by at least 11 players, it is closed off so that players are forced to select a different board. Any open board can be selected for play. Also, once each player has completed a particular board, they are not allowed to play it again.

## Purpose

The purpose of The Cure is to translate the knowledge of the players along with their ability to process textual information into a ranked list of genes for use in the development of predictors for breast cancer prognosis. This translation is enacted when the players select genes in the game. We record the gene selections and apply aggregation functions to produce gene rankings that reflect the consensus knowledge of the player community.

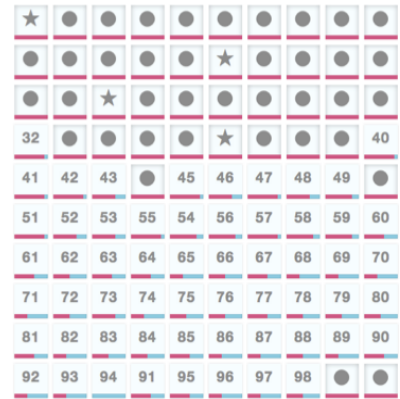

**Fig. 3.** Board selection view. Stars indicate boards the active player has completed, circles indicate boards that have been completed by a sufficient number of different players, and numbers indicate open boards. The pink progress bar indicates how close the community is to finishing the board.

# 3  RESULTS

The results presented here are derived from games played between September 7, 2012 and September 5, 2013. In that time 1077 player accounts were created and a total of 15669 games were played (including training games). 9904 games were played on the cancer datasets.

## 3.1  Players and Games Played

Based on the self-reported data collected during registration, the player population was mixed in terms of education, orientation as a biologist, and declared knowledge of cancer. In total, 35% of the players had a graduate degree, 29% had an undergraduate degree, and 36% did not declare any degree. 32% of players considered themselves biologists while 63% did not, with 5% not responding. 33% of players declared that they were knowledgeable about cancer biology, 60% did not, and 7% declined to respond.

Over the course of the year, the number and demographics of players registering to play per month fluctuated (Fig. 4). In the first two months, 36 and 37% of the players who registered had PhDs. After those months, the percentage for the next four months dropped to around 15% and then fluctuated between 5 and 10% thereafter. We observed two notable spikes in player registrations. The first coincided with the launch of the game, its presentation at Genome Informatics 2012 and its advertisement as part of the Sage Bionetworks DREAM7 breast cancer prognosis challenge. The second, in May of 2013, is likely related to a posting on the popular website i09 (http://io9.com/these-cool-games-let-you-do-real-life-science-486173006) which occurred on May 1, 2013.

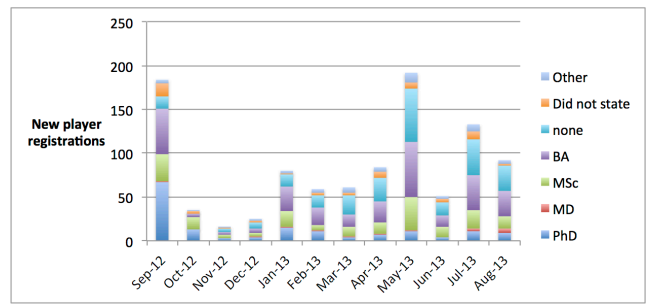

**Fig. 4.** New player registrations per month, with degree.

The total number of games played roughly followed the trends observed for new player registrations. The most games played on a single day was 550, on May 2, immediately after the i09 posting.

The number of games played per player followed a power law consistent with most studies of the quantity of voluntary contributions in open environments (e.g. Wikipedia contributions) (Javanmardi, et al., 2009). 243 players played more than 10 games, 28 players played more than 100 games and the most prolific player ('oneoff64') played 718 games (Fig. 5).

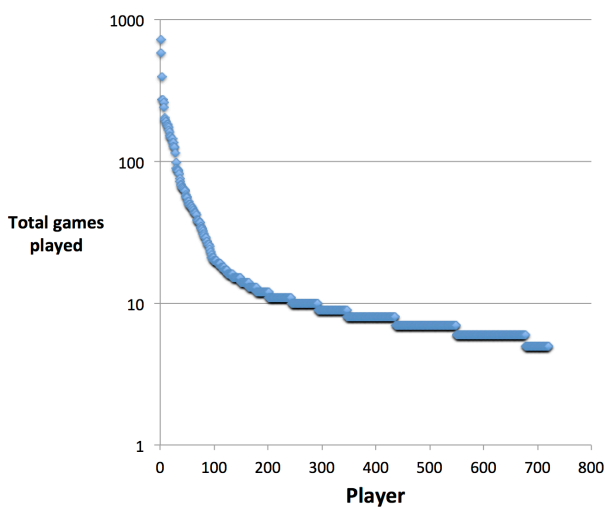

**Fig. 5**. Games played per player.

### 3.2 Aggregation function for gene ranking

Each time a player selects a gene in a game, they are indicating to the system their intuition of that gene's relevance for predicting breast cancer survival. That intuition may be based on their knowledge, on inferences drawn from gene annotation information or solely on random speculation. By aggregating the data collected from many different players across many different games, we tried to eliminate the noise from random clicking and reveal the community consensus with regard to predictive genes.

Given a set of recorded games, our gene ranking method is as follows. For each gene $g$ we estimate the frequency of selection $F(g)$ as:

$$F(g) = \frac{S(g)}{O(g)}$$

$O(g)$ equals the number of times the gene $g$ appeared in a played game. Some genes appear on multiple boards, all boards are played by multiple players, all occurrences are counted. $S(g)$ is the number of times the gene was selected by the human player.

We then empirically calculated a one-tailed p value for each value of $F$ given $O$ through simulations of random game play. The p values indicate the chances of observing a value of $S$ or greater given $O$ assuming that all gene selections were random. Importantly, they allow for comparisons between genes with different numbers of occurrences. For example, the known apoptosis regulator BCL2 gene occurred in 13 played games ($O = 13$) and was selected in 10 of those games ($S = 10$), thus $F$ for BCL2 was 0.77 with p < 0.0001. Our simulations stopped at 10,000 iterations per value of $O$, hence p values below 0.0001 cannot be reported. On the other end of the spectrum, the AARD gene (of unknown function) appeared in 33 played games ($O = 33$), was selected 3 times ($S = 3$), had an $F$ of 0.09 with p = 0.91. Given any collection of played games, we generate gene rankings based on the estimated p values for each value of $F$. We can thus assemble gene sets based on different groups of games as well as different p value cutoffs.

### 3.3 Gene set evaluations

We evaluated three game collections: '**all**', '**expert**', and '**inexperienced**'. 'All' considers games from all players, 'expert' includes games from players that indicated that they had either a Ph.D. or an M.D. and knowledge of cancer, and 'inexperienced' includes just the games played by people without an advanced degree, with no knowledge of cancer and that were not biologists. Only the first game per player per board is used for the analysis to reduce the chances of individual players essentially over-fitting the training data for a particular board. Each game should reflect only the player's thoughts about the best genes for that board prior to seeing the results of the decision tree analysis.

For all the results reported here, we select genes with p <= 0.001. At that threshold, we observed 61 genes in the 'all' group, 85 in the 'expert' group and 13 in the 'inexperienced' group (Table 1).

**Table 1.** Predictor gene sets derived from The Cure.

| Player group | N genes, p <=0.001 | Games considered | Contributing players |
|---|---|---|---|
| All | 61 | 4314 | 477 |
| Expert | 85 | 1106 | 52 |
| Inexperienced | 13 | 1643 | 231 |

One gene, CASP1, appeared in all three sets. The 'all' gene set included 35 genes that also appeared in the expert set as well as 4 genes from the inexperienced set (Fig. 6). Aside from CASP1, there was no overlap between the expert and inexperienced gene sets.

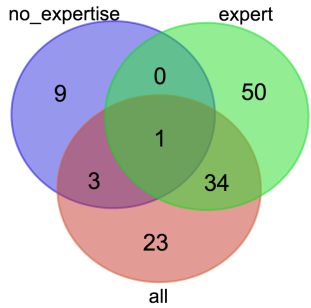

**Fig. 6.** Overlap of game-derived gene sets.

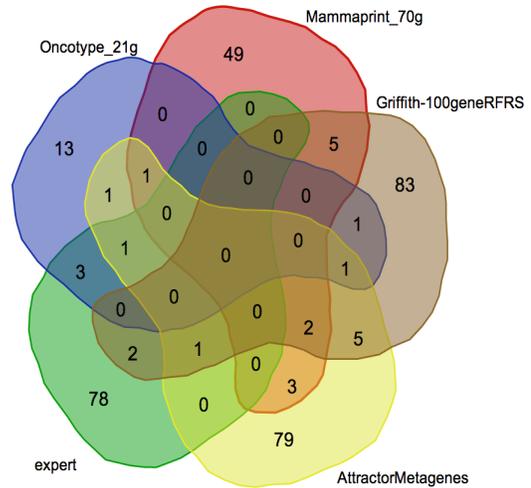

**Fig. 7**. Overlap of 'expert' gene set derived from game data (in green) with prior published predictor gene sets.

**Enrichment analysis of gene sets**

Given the gene sets identified above, the next question is whether or not they are relevant to breast cancer. Has knowledge successfully been transferred from the player population into the game?

Using WebGestalt (Wang, et al., 2013), we found both the 'expert' gene set and the 'all' gene sets to be significantly enriched for cancer-related diseases while the 'inexperienced' set was not significantly enriched for any diseases (Table 2). The background genes used for the enrichment analysis statistics corresponded to the 3731 genes that appeared in at least one game. The disease term with the most significant corrected p value in both the 'all' and 'expert' gene sets was 'cancer or viral infections'. All of the top ten disease terms for both gene sets correspond to various kinds of cancer or cancer processes such as 'recurrence' and 'disease progression'. Though they do not appear in the top ten results, "Breast neoplasms' and 'Carcinoma, Ductal, Breast' are significantly represented in both gene sets (p < e-05).

**Table 2.** The top ten most significantly enriched disease terms for the 'expert' gene set. WebGestalt enrichment statistics are also shown for the 'all' gene set for the same terms.

| Disease term | Expert players (85 genes) | | All (61 genes) | |
|---|---|---|---|---|
| | Genes in set | Adj. P value (BH) | Genes in set | Adj. P value (BH) |
| Cancer or Viral Infections | 37 | 5.5e-16 | 25 | 8.1e-10 |
| Neoplasms | 32 | 4.7e-13 | 22 | 1.6e-08 |
| Urogenital Neoplasms | 23 | 2.7e-11 | 12 | 9.0e-05 |
| Cell Transformation, Neoplastic | 16 | 4.7e-11 | 13 | 3.0e-09 |
| Stomach Neoplasms | 14 | 2.6e-08 | 8 | 2.0e-04 |
| Disease Progression | 16 | 3.7e-08 | 13 | 2.5e-07 |
| Neoplastic Processes | 20 | 5.1e-08 | 18 | 1.3e-08 |
| Recurrence | 14 | 5.1e-08 | 11 | 1.1e-06 |
| Intestinal Neoplasms | 15 | 6.1e-08 | 6 | 0.01 |
| Necrosis | 15 | 1.1e-07 | 13 | 1.8e-07 |

**Comparison to established predictor gene sets**

In addition to the disease enrichment analysis, we measured the overlap between the game-derived gene sets and predictor gene sets used in commercial prognostic tests and from recent publications. Figure 7 shows the overlaps between the expert game gene set, the 21 genes used in the OncotypeDx test (Paik, 2007), the 70 genes in the Mammaprint test (van 't Veer, et al., 2002), 100 genes recently identified via Random Forest analysis (RFRS) (Griffith, et al., 2013), and 94 genes recently identified via the Attractor Metagenes approach (Cheng, et al., 2013b). Genes in the gold standard sets that never appeared in a played game were removed from the comparison. (e.g., only 58 of the 70 genes in the Mammaprint set were used.) The 'expert' gene set contained four of the OncoType genes, zero of the Mammaprint genes, three of the RFRS genes and two of the Attractor Metagenes. Based on a Fisher's exact test, there was a statistically significant overlap with only the OncoType genes (p = 2.026E-4).

**Classifier evaluations**

The gene set comparisons and enrichment analyses described above show clearly that the gene sets generated from the game data are non-random with a significant representation of genes that are related to cancer. The final question addressed here is how well the game-derived gene sets do when used to create classifiers for predicting breast cancer survival.

We conducted two experiments, each involving the development of machine learning models for predicting 10-year survival based only on gene expression information. In the first, we trained a support vector machine (SVM) classifier using gene expression data from the Metabric dataset (Curtis, et al., 2012) and tested it on the Oslo validation dataset generated for the Sage Dream7 breast cancer challenge (Margolin, et al., 2013). In the second, we used the dataset from (Griffith, et al., 2013), using the same division of training/test data described in that publication. In both cases, we varied the gene sets provided to the classifiers and measured the performance of each gene set based on the accuracy of the SVM on the samples in the corresponding test set. As shown in Figure 8, both the 'expert' and 'all' gene sets from the game performed comparably to the OncoType, Mammaprint, RFRS, Attractor MetaGenes, and to gene sets selected in an expert literature review (Lauss, et al., 2008). In fact, the 'expert' gene set from The Cure had the highest accuracy on the Griffith test set and the third highest accuracy on the Oslo test set.

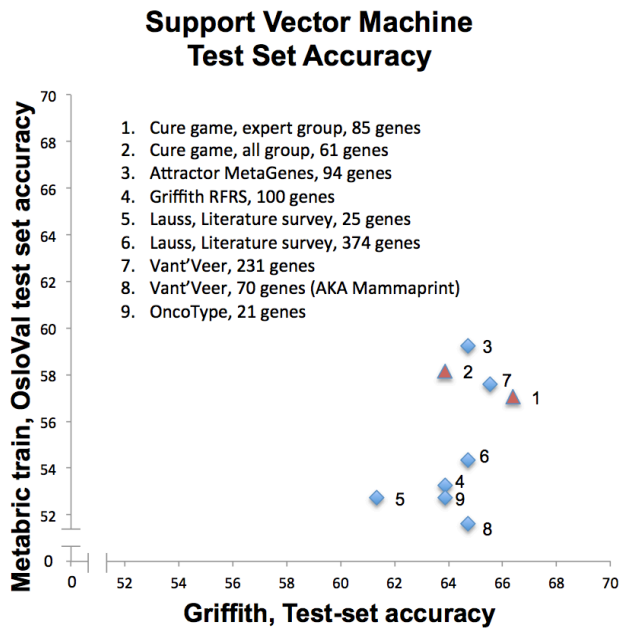

**Fig. 8**. Evaluation of accuracy of models trained to predict ten-year survival using gene sets derived from the game and prior gene sets from the breast cancer literature.

Based on these experiments and others employing different machine learning methods (data not reported) we could not establish a statistically significant difference between the performance of models trained using the game-derived gene sets versus models trained with gene sets from more established methodologies. While we could not prove that the game-derived gene set was better than the other gene sets in a statistically significant manner, none of the other gene sets – including those used in commercial tests – were found to be consistently better either. The lack of a clear 'winner' in this analysis reinforces the concept that there are actually many different gene sets whose expression signatures are nearly equally predictive of breast cancer prognosis (Venet, et al., 2011). Identifying the optimal combination of genes, clinical features (e.g. age, lymph node status) and machine learning approach remains a future challenge.

### 3.4 Player survey

The Cure managed to attract and engage a surprisingly large number of people. To ascertain more about the player population we conducted a survey of registered players as of November 2013. We sent an email to the 1162 players who had entered an email address when they registered inviting them to answer questions about themselves, their motivations for playing and their experience with the game. We received responses from 119 participants. While the respondents represent only about 10% of the total player population (and likely a more motivated segment), the responses do provide some interesting insights.

The first and perhaps most telling question in the survey was: "Why did you sign up for The Cure?" Overall, 71% indicated that they played to help breast cancer research, 52% played to learn something and just 44% played in order to have fun. Respondents could select multiple answers. Given the design of The Cure website ("Play Games, Cure Cancer!") as well as the way it was promoted, it is surprising to see that the game aspect was actually the least motivational of the three. While we feel that developing this system as a game had a strong positive effect on recruitment and engagement, its clear that there is a large pool of people that are highly motivated to contribute to breast cancer research in any way they can. The game was simply one more vehicle through which they could try to help. In some cases, this motivation is likely very personal; 64% of respondents indicated that they know or have known someone that has or has had breast cancer.

Looking at the player demographics, we found that 59% of the respondents were male and 41% female. The largest age brackets were 21-29 (34.5%) and 30-39 (28.6%) (Fig. 9).

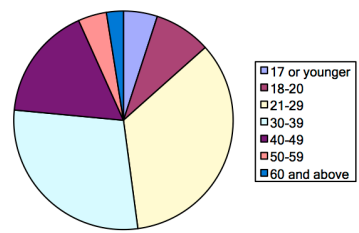

**Fig. 9**. Ages of players.

Expanding on the expertise information collected when players registered, we asked players to categorize their knowledge of breast cancer. The most popular answer by a wide margin was the middle expertise level "I know some biology and have some understanding of what cancer is" at 57.1% with numbers decreasing towards the high and low expertise levels (Fig. 10).

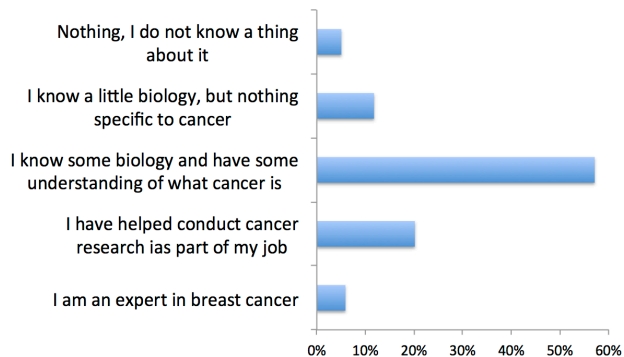

**Fig. 10.** Levels of breast cancer knowledge among players.

Finally we asked players whether the game was fun and whether or not they learned anything. Most (66%) found the game to be "A little bit entertaining", 14% found it to be "very fun", and 19% found it "not at all" fun. The results for learning are similar, with 62% feeling that they "learned a little bit", 9% that they "learned a lot", and 29% that they "did not learn thing".

In summary, the survey showed that The Cure reached a demographically diverse audience containing both experts and novices, that most players found the game mildly entertaining and educational, and that the dominant motivation for playing was to help breast cancer research.

## 4   DISCUSSION

The principal contribution of this work is to show that crowdsourcing games can be developed as a means to address problems that require expert-level knowledge. While previous work on games such as Foldit and citizen science projects like Galaxy Zoo have focused on visual problems that do not require any knowledge on the part of the participant, the task presented in The Cure was knowledge intensive. In order to successfully participate, players either had to bring significant prior experience or be willing to invest a substantial amount of time learning. The numbers of players and games played provide evidence that there is a large audience of experienced and curious people that are willing to participate in knowledge and reasoning-intensive challenges. The analysis of the gene sets produced from the aggregated data shows that the game successfully acted as a vehicle for capturing knowledge from this newly formed community. While this game did not generate a better predictor of breast cancer prognosis, the results are in general very encouraging.

**Limitations**

The game described here was an early-stage prototype with many limitations in terms of both its ability to achieve its purpose and its ability to entertain players. Chief among the former was that the pre-built boards severely constrained the number of different feature combinations that players could explore. The vast majority of possible gene sets simply could never be examined within this game framework. Further, because the aggregation function ranked individual genes rather than gene sets, it is unlikely that it would identify optimal feature combinations. In future iterations, it would be beneficial to adapt the game to allow advanced players more freedom to explore the feature space while still maintaining the competitive dynamics that made the game entertaining. In addition to providing greater room for exploration, the game could be made much more fun overall – especially for not-yet-expert players. The current formulation was highly repetitive and had an extremely steep learning curve. The transition from the brief training stage to the real games was abrupt and left many players confused. In the future, both the fun factor and the learning aspects of the game could be improved by implementing different levels of difficulty, providing more educational information in the early stages, and diversifying the tasks presented to players. The code for The Cure game is open source and we would warmly welcome any contributions or adaptations (https://bitbucket.org/sulab/thecure/).

## 5   CONCLUSION

There is a large, heterogeneous population of people on the Internet that actively seek ways to use their minds to help solve important problems. Games such as The Cure provide one avenue to tap into this hidden resource for biomedical discovery.

## 6   METHODS

**Composing gene sets for four rounds of game play**

The game play data presented here was collected in four distinct rounds, with each round consisting of a set of 100 boards (sets of 25 distinct genes). We chose gene sets for each round by first identifying a list of 2500 'interesting' genes based on unsupervised analysis of a genomic dataset. We then sampled randomly from this gene list to produce the boards for the game.

**Dataset used for rounds 1 and 2 (Sage DREAM7 challenge)**

The first two rounds corresponded to the two iterations of the training data provided for the Sage DREAM7 challenge (Margolin, et al., 2013). In both cases Sage provided a processed subset of the METABRIC dataset (Curtis, et al., 2012) with information about gene expression, copy number variation (CNV), and clinical features including survival information. We used the survival data to group the samples into two classes: those with less than ten-year survival from the point of diagnosis and those with greater than ten-year survival. For these data sets we selected genes to include in the game based on both gene expression information and CNV data. First we used the CNV information to rank the genes based on the sum of squares across all samples and selected the top 1000 genes. Next, we developed a ranking of the genes based on expression data as follows:

(1) Remove probes with poor matches to genes based on the ReMOAT annotation data (Barbosa-Morais, et al., 2010).
(2) Remove probe sets where the maximum expression value across all samples is lower than the overall median
(3) Rank probe sets based on their variance across all samples

Finally, we merged the genes selected based on CNV information with the expression-based ranking to produce a single table with 2500 unique genes.

**Dataset used for rounds 3 and 4 (Griffith dataset)**

Rounds three and four used the meta gene expression dataset assembled for (Griffith, et al., 2013). For this dataset, genes were selected following the same approach as described in the original publication (where at least 20% of samples should have intensities greater than the background threshold and the coefficient of variation is between 0.7 and 10). Genes that passed those filters were then ranked based on their variance across all samples to produce a set of 2500 unique genes.

**Composing boards**

In the first round, we seeded each board with one gene with a high ReliefF (Kononenko, 1994) value based on the combined CNV and expression data with all other genes selected randomly from the set of 2500. (Note that players did not preferentially select the seeded genes.) In the three subsequent rounds, we first created 50 boards by randomly sampling from the 2500. Then we selected the second set of 50 boards by sampling from the genes used in the first 50 boards such that each gene appears in two boards per round. While reducing the coverage of genes, this strategy allowed players to assess each gene in multiple contexts, hopefully allowing a fairer assessment of the gene's overall value. Each round had some overlap in terms of genes used. In total, 3,731 distinct genes were used in boards played in the game.

**Scoring hands**

Each time a gene was added to a player's hand during a game, the server immediately responded with a score for the genes in their hand and a decision tree inferred using those genes. This was accomplished using source code adapted from the Waikato Environment for Knowledge Analysis (WEKA) (Witten, et al., 1999) operating in a Tomcat Web server. We used WEKA's implementation of the C4.5 decision tree induction algorithm as well as its code for cross-validation. The scores were accuracy estimates from 10-fold cross-validation experiments and the trees were inferred using the provided genes and all training instances. Trees were constructed based on both CNV and expression data for round 1, and just expression data for rounds 2-4.

**Classifier evaluations**

We used WEKA's default implementation of the sequential minimal optimization algorithm for training the SVMs used in the gene set analysis.

## ACKNOWLEDGEMENTS

We would like to thank all of the people who played The Cure.

*Funding*: Research reported in this publication was supported by the National Institute of General Medical Sciences of the National Institutes of Health under award numbers R01GM089820 and R01GM083924, and by the National Center for Advancing Translational Sciences of the National Institute of Health under award number UL1TR001114.